\documentclass[a4paper]{article}
\usepackage{enumerate}
\usepackage[ansinew]{inputenc}
\usepackage[T1]{fontenc}
\usepackage{amsmath,amsthm}
\usepackage{amssymb}
\usepackage{empheq}

\newcommand{\beq}{\begin{equation}}
\newcommand{\eeq}{\end{equation}}

\title{ \vspace{-100pt} \hfill \hbox{\small{DIAS-STP-21-19}} \vspace{100pt}\\
  Constrained Dynamics in the Hamiltonian formalism}

\author{Brian P. Dolan\footnote{email: bdolan@stp.dias..ie} \\ \\
  \textit{\small School of Theoretical Physics}\\
  \textit{\small Dublin Institute for Advanced Studies}\\
  \textit{\small 10 Burlington Rd., Dublin, Co. Dublin, Ireland}}

\begin{document}

\maketitle

\abstract{These are pedagogical notes on the Hamiltonian formulation of constrained dynamical systems. All the examples are finite dimensional, field theories are not covered, and the notes could be used by students for a preliminary study before the infinite dimensional phase space of field theory is tackled.

  Holonomic constraints in configuration space are considered first and Dirac brackets introduced for such systems. It is shown that Dirac brackets are a projection of Poisson brackets onto the constrained phase space and the projection operator is constructed explicitly.  More general constraints on phase space are then considered and exemplified by a particle in a strong magnetic field. First class constraints on phases are introduced using the example of motion on the complex projective space ${\mathbf{C P}}^{n-1}$. Motion of a relativistic particle in Minkowski space with a reparameterisation invariant world-line is also discussed.

  These notes are based on a short lecture course given at Bhubaneswar Indian Institute of Technology in November 2021.

}

\tableofcontents

\section{Introduction}
These notes are on constrained dynamics of Hamiltonian systems in classical mechanics. This is a class of problems that is particularly important as a preliminary step in understanding the quantum version of such systems, especially in
field theory --- gauge theories and general relativity in particular.
This motivated Dirac to study this problem in the 1960s and there is now a number of articles explaining the concepts.  But many such discussions (especially
of first class constraints, which are defined later)
go straight to field theory, though Dirac's formalism is of course equally applicable to finite dimensional systems which  are simpler than field theories, as field theories have an infinite number of degrees of freedom.  These notes are written to explain the basic concepts in as simple a manner as possible, using only a finite number of degrees of freedom.

There is a number of good books on the infinite dimensional case, most notable by Dirac \cite{Dirac}, but see also \cite{HT}.  It is hoped that these notes will offer the student a good foundation to go on and study the field theory case in these references.

\section{Summary of Lagrangian and Hamiltonian formulations}

In the Lagrangian formulation of classical mechanics one introduces generalised co-ordinates $q^a(t)$ in configuration space, $a=1,\ldots,N$,
and obtains equations of motion for the $q^a$ by applying a variational principle to the action
\[S[q(t_1), q(t_2)]=\int_1^2 L (q,\dot q)dt,\]
with the end-points $q(t_1)$ and $q(t_2)$ fixed. 
\begin{align}
  \delta S =0    \qquad 
  &\Rightarrow \qquad \int_1^2 \left( \frac{\delta L}{\delta \dot q^a}\delta \dot q^a
                       + \frac{\delta L}{\delta q^a}\delta q^a\right)dt  =0 \nonumber\\
  &\Rightarrow \qquad \int_1^2 \left(-\frac{d}{d t}\left( \frac{\delta L}{\delta \dot q^a}\right)
    + \frac{\delta L}{\delta q^a}\right)\delta q^a dt  =0 \nonumber
\end{align}
where we have integrated by parts  with the end-points fixed, $\delta q(t_1)=\delta q(t_2)=0$. 
Since the variations $\delta q^a$ are otherwise arbitrary we obtain the Lagrangian equations of motion
  \beq\frac{d}{d t} \left( \frac{\partial L}{\partial \dot q^a} \right) = \frac{\partial L}{\partial q^a}=0.\label{Leom}\eeq

In the Hamiltonian formalism one first derives the canonical momenta
\beq p_a(q,\dot q) = \frac{\partial L}{\partial \dot q^a}\label{piqdot}\eeq
and the Hamiltonian is defined to be\footnote{We are using the Einstein summation convention,
  $\dot q^a \dot q^a := \sum_{i=1}^N \dot q^a \dot q^a$  -- repeated indices are summed over.}
\beq H(q,p) = p_a \dot q^a -L(q,\dot q)\label{H}.\eeq
The Hamiltonian $H(q,p)$ is a function on the $2N$-dimensional phase space
with co-ordinates $(q^a,p_a)$ and in order to obtain $H(q,p)$ it is necessary to invert (\ref{piqdot}) and express  $\dot q^a(q,p)$ as a function
of $q^a$ and $p_a$. 
Assuming this can be done a variation of (\ref{H}) gives
\begin{align*}
  \delta H &= \frac{\partial  H}{\partial q^a} \delta q^a  + \frac{\partial  H}{\partial p_a} \delta p_a \\
  &=  (\delta p_a) \dot q^a + p_a (\delta \dot q^a)  - \left(\frac{\partial L}{\partial \dot q^a}\right) \delta \dot q^a  - \left(\frac{\partial L}{\partial q^a}\right) \delta q^a\\
           &= \dot q^a (\delta p_a)  - \left(\frac{d p_a}{d t} \right) \delta q^a
\end{align*}
from the definition (\ref{piqdot}) and the Lagrangian equations of motion (\ref{Leom}).
Equating co-efficients the Lagrangian equations of motion are equivalent to the Hamiltonian equations of motion
\beq \frac{d q^a}{d t} = \frac{\partial H}{\partial p_a}, \qquad \frac{d p_a}{d t} = -\frac{\partial H}{\partial q^a}.\label{Heom}\eeq
The time evolution of any function $F(q,p)$ on phase space is given by\footnote{For simplicity we assume that $F(q,p)$ has no explicit time dependence.
  If there is explicit as well as implicit time dependence, $F(q,p,t)$, then
\[\frac {d F}{d t}=  \frac{\partial F}{\partial t}+ \{F,H \}. \] }
\[ \frac{d F}{d t} =
  \frac{\partial F}{\partial q^a}\dot q^a
  +  \frac{\partial F}{\partial p_a}\dot p_a
  =  \frac{\partial F}{\partial q^a}\frac{\partial H}{\partial p_a}
  -  \frac{\partial F}{\partial p_a} \frac{\partial H}{\partial q^a}
=\{F,H\} \]
where the \textit{Poisson bracket} of any two functions $F$ and $G$ on phase space is defined to be
\[ \{F,G \} =  \frac{\partial F}{\partial q^a}\frac{\partial G}{\partial p_a}
  -  \frac{\partial F}{\partial p_a} \frac{\partial G}{\partial q^a}\]
so
\[ \frac{d F}{d t} = \{ F,H\}.\]

Poisson brackets satisfy the following identities, for any three functions,
$F$, $G$ and $K$, on phase space
\begin{align}
  \{F, G\}&=-\{G, F\}\\
  \{F+G, K\}&=\{F, K\} +\{G, K\}\\
  \{F G, K\}&=F \{G, K\} +  \{F, K\} G\\
  \{F,\{G, K\}\} &+\{G,\{K, F\}\}+\{K,\{F, G\}\}=0. \label{Jacobi}
\end{align}
Equation (\ref{Jacobi}) is known as the \textit{Jacobi identity}.

\section{Holonomic constraints on configuration space}

The imposition of constraints on configuration space in the Lagrangian formalism is in principle straightforward.
Suppose we want to constrain the $q^a$ to sit on an $(N-1)$-dimensional  hypersurface in configuration space described by an
equation $\phi(q)=0$. For example if $q^a = x^a$ are co-ordinates in $N$-dimensional Euclidean space, $\phi(x)=x^a x^a - R^2$ constrains the motion to lie on a $(N-1)$-dimensional sphere of radius $R$.
The constraint imposes conditions on how the $q^a$ can vary,
\[ \delta \phi =0 \quad \Rightarrow \quad (\partial_a \phi) \delta q^a=0\]
and this is called a \textit{holonomic} constraint. A more general, non-holonomic, constraint on the possible variations
on configuration space could be something like
\[ \Phi_a(q) \delta q^a=0\]
where the $N$-functions $\Phi_a(q)$ are not derivatives of a single function.\footnote{For example\[ q^2 \delta q^1 +  q^1 \delta q^2 =0\]  is a holonomic constraint, with $\phi = q^1 q^2$, while \[ q^2 \delta q^1 -  q^1 \delta q^2 =0\] is non-holonomic, there is no function $\phi(q)$ for which $q^2 \delta q^1 -  q^1 \delta q^2=\frac{\partial \phi}{\partial q^a} \delta q^a$.}
In this section we shall focus on holonomic constraints on configuration space.

In Cartesian co-ordinates let
\beq L = \frac{m}{2} \dot x^a \dot x^a - { V }(x)\label{FreeL}\eeq
be the Lagrangian for a particle of mass $m$ moving under the influence of a potential ${ V }(x)$.
There are two ways to implement the constraint:

\begin{enumerate}
\item We can solve the constraint $\phi(x) =0$ to eliminate one of the $x^a$,
  e.g. $\phi(x)=0$ gives $x^N(x^1,\ldots,x^{N-1})$ explicitly as a function of $(x^1,\ldots,x^{N-1})$,
  reducing configuration space to $N-1$ dimensions.
\item  We can increase configuration space to $( N+1)$ dimensions by introducing a new generalised co-ordinate $q^{N+1}=\lambda$
and modifying the Lagrangian to
\beq {\widetilde L} = \frac{m}{2} \dot x^a \dot x^a - { V }(x) - \lambda \phi(x).\label{Llambda}\eeq
Now $\frac{\partial {\widetilde L}}{\partial \dot \lambda}=0$, so
$\frac{d}{d t} \bigl(\frac{\partial  {\widetilde L}}{\partial \lambda}\bigr)=0$,  and the Lagrange equation of motion for $\lambda$ reduces to
\beq \frac{\partial  {\widetilde L}}{\partial \lambda} = \phi(x) =0,\label{constraint}\eeq
thus imposing the constraint as another equation of motion.  $\lambda$ is called a \textit{Lagrange multiplier} for the constraint $\phi(x)=0$.
There are now $(N+1)$ degrees of freedom and $(N+1)$ equations of motion, including (\ref{constraint}), so the problem is well determined. In particular
$\lambda(x,\dot x)$ will in general be a specific function of $x^a(t)$ and $\dot x^a(t)$ when the equations of motion are solved. 
\end{enumerate}

In this section we shall develop the Hamiltonian formalism associated with such a holonomic constraint. The Hamiltonian associated with the Lagrangian (\ref{FreeL}) is
\[ H(x,p) = \frac{1}{2m} p_a p_a + { V }(x).\]
Including the Lagrange multiplier for the constraint the na{\"\i}ve Hamiltonian is 
\[ \widetilde H(x,p,\lambda) = \dot x^a p_a - {\widetilde L} = \frac{1}{2 m} p_a p_a + { V }(x) +\lambda \phi(x)=H(x,p)+\lambda \phi(x),\]
since $p_\lambda = \frac{\partial {\widetilde L}}{\partial \dot \lambda}=0$.

Now we have a conundrum: the Hamiltonian equation of motion for $\lambda$ is
\[ \dot \lambda =\frac{\partial \widetilde H}{\partial p_\lambda}=0,
\]
so $\lambda$ is a constant for any potential ${ V }(x)$, but the Lagrangian analysis gives $\lambda(x,\dot x)$ a function of $x^a$ and $\dot x^a$ and there will certainly exist potentials for which this is not a constant.
One way out of this apparent paradox is not to treat $\lambda$ as a dynamical
variable in the Hamiltonian but instead consider a modified Hamiltonian 
\[H_T(x,p) = p_a \dot x^a - L + u(x,p) \phi(x) = H(x,p) + u(x,p)\phi(x)\]
($T$ for ``total'') where $u(x,p)$ is an unknown function on $2N$-dimensional phase space.
On the constraint surface $H_T=H$ but in the Hamiltonian formulation $\delta x^a$ and $\delta p_a$ are independent variations, they are not constrained.

The Hamiltonian equations of motion arising from $H_T$ are
  \begin{align*}
\dot p_a  & =  -\frac{\partial H_T}{\partial x^a}=-\frac{\partial H}{\partial x^a} - u \frac{\partial \phi}{\partial x^a} -\frac{\partial u}{\partial x^a}\phi,\\
\dot x^a  & = \ \frac{\partial H_T}{\partial p_a}=\frac{\partial H}{\partial p_a} + \frac{\partial u}{\partial p_a}\phi.    \end{align*}
    
On the constraint surface $\phi=0$ and these reduce to 
  \begin{align*}
\dot p_a  & =-\frac{\partial H}{\partial x^a} - u \frac{\partial \phi}{\partial x^a},\\
    \dot x^a  & =\frac{\partial H}{\partial p_a}    \end{align*}
$-u \frac{\partial \phi}{\partial x^a}$ can be interpreted as a force normal to the surface that keeps the motion on the surface (the gradient $\nabla \phi$ is always normal to the constraint surface).
  
  A convenient notation is to use ``$\approx$'' to denote equations that are only true when the constraint is applied and reserve ``$=$'' for equations that are always true, without the constraint. We shall call equations involving ``$\approx$'' \textit{weak} equations and 
  equations involving ``$=$'' \textit{strong} equations. Thus $\phi(x)\approx 0$ and $H_T \approx H$, but $\phi(x) \ne 0$ and $H_T \ne H$.
  We shall call $\phi(x)\approx 0$ a \textit{primary constraint}.
  
  With this notation the equation of motion are
   \begin{align*}
\dot p_a  & \approx-\frac{\partial H}{\partial x^a} - u \frac{\partial \phi}{\partial x^a},\\
    \dot x^a  & \approx\frac{\partial H}{\partial p_a}.    \end{align*}
In the following examples we shall see that the time evolution constrains the from of $u(x,p)$.

  \subsection{$\phi(x)=x^N -c_0$ \label{Plane}}
  The constraint $x^N=c_0$ is particularly simple and is trivial to implement in the Lagrangian formalism, but it is analysed here in the Hamiltonian formalism to illustrate a technique which is useful in more complicated situations.

  The constraint $\phi(x)=x^N - c_0$, with $c_0$ a constant,  forces $x^N$ to be a constant, but this must be true for all times so we must check the time evolution of the constraint.  With
  \[  H_T(x,p) = \frac{1}{2 m} p_a p_a +{ V }(x) + u \phi(x)\]
  we have
  \[ \dot \phi = \dot x^N=\{\phi,H_T\}= \frac{\partial \phi}{\partial x^a}\frac{\partial H_T}{\partial p_a} - \frac{\partial \phi}{\partial p_a}\frac{\partial H_T}{\partial x^a}=\frac{p_N}{2 m}. 
  \]
  If $x^N$ is a constant then it must be the case that $\dot x^N \approx 0$ is weakly zero, hence we get another constraint
  \[\chi=p_N\approx 0,\]
  this is called a \textit{secondary constraint}.
  We must further check the time evolution of $\chi$,
  \[ \dot\chi =-\frac{\partial H_T}{\partial x^N}
    =-\frac{ \partial { V }}{\partial x^N} -u-\frac{\partial u}{\partial x^N} (x^N-c_0),\]
  and $\dot \chi \approx 0$ requires that $u\approx -\frac{ \partial { V }}{\partial x^N}$.  If we define
  \[ U=-\left.\frac{\partial { V }}{\partial x^N}\right|_{x^N=c_0}\]
  then $u\approx U$. The time evolution is consistent provided we set $u\approx U$.
  
  But we now have an apparent paradox,
  \[x^N \approx \mbox{const} \qquad \mbox{and} \qquad p^N \approx 0,\]
  yet, by definition, the Poisson bracket $\{ x^N,p_N\}=1$.  The constraints are not compatible with the Poisson bracket structure (note that all derivatives and Poisson brackets must be calculated \textit{before} the constraints are applied). We can save the situation by modifying the Poisson bracket.
  First we adopt a more convenient notation, let
  $\phi_1=\phi$ and $\phi_2=\chi$. Then construct the $2\times 2$
  anti-symmetric matrix
  \[ C_{I J} = \{\phi_I,\phi_J\}\]
  where $I,J=1,2$.  Explicitly, in this example,
  \[ C= \begin{pmatrix} 0&1\\ -1&0 \end{pmatrix}
    \qquad \mbox{and}\qquad  C^{-1} = \begin{pmatrix} 0&-1 \\ 1&0 \end{pmatrix}.\]
  Now define a modified Poisson bracket, called the \textit{Dirac bracket}, by
  \beq\{ F, G\}_* = \{F, G\} - \{F,\phi_I\} \bigl(C^{-1}\bigr)^{I J} \{ \phi_J, G\}. \label{DiracBracket}\eeq
  Since
  \[ \{F, \phi_1\} = -\frac{\partial F}{\partial p_N},
    \quad \{F, \phi_2\} = \frac{\partial F}{\partial x^N},
    \quad \{\phi_1, G\} = \frac{\partial G}{\partial p_N},
    \quad \{\phi_2, G\} = -\frac{\partial G}{\partial x^N} \]
  this is
  \begin{align*}
    \{ F, G\}_* &= \{F, G\}
    -\begin{pmatrix}-\frac{\partial F}{\partial p_N},& \frac{\partial F}{\partial x^N}\end{pmatrix}
    \begin{pmatrix} 0&-1 \\ 1&0 \end{pmatrix}
                               \begin{pmatrix}
                                 \frac{\partial G}{\partial p_N}\\ -\frac{\partial G}{\partial x^N} \end{pmatrix}\\
    &=\{F, G\} - \left(-\frac{\partial F}{\partial p_N}\frac{\partial G}{\partial x^N} + \frac{\partial F}{\partial x^N}\frac{\partial G}{\partial p_N}\right).
    \end{align*}
    In particular
 \begin{align*}
      \{ x^a,x^b\}_*&=0\\
      \{ x^a,p_b\}_*&= \delta^a{}_b\\
      \{ p_a,p_b\}_*&=0\\
      \{x^N,p_a\}_*&=\{x^a,p_N\}_*=0\\
   \{x^N,p_N\}_*&=0,    
      \end{align*}
    with $a,b=1,\ldots N-1$.
    These are strong, not weak, equations and as such it does not matter whether we apply the constraints before or after calculating the
    Dirac brackets: $x^N\approx c_0$ and $p_N\approx 0$ are perfectly compatible with $\{x^N,p_N\}_*=0$.
    
    We can now take any two functions $F(x^1 \kern -2pt ,\ldots ,x^N\kern -3pt ,p_1,\ldots,p_N)$ and \break
    $G(x^1\kern -2pt,\ldots,x^N\kern -3pt,p_1,\ldots,p_N)$ on the $2N$-dimensional phase space and calculate the Dirac bracket, $\{F,G\}_*$.  The conditions\footnote{We no longer need to distinguish between strong and weak equalities.}  $x^N=c_0$ and $p_N=0$ can be imposed either before or after calculating the Dirac brackets the answer will be the same, we just eliminate $x^N$ and $p_N$, by setting $x^N=c_0$ and $p_N=0$, and end up with the $(2N-2)$-dimensional phase space with co-ordinates $(x^1,\ldots,x^{N-1},p_1,\ldots,p_{N-1})$.

    This is a particularly simple example, we could just as easily have set $x^N=c_0$ and $p_N=0$ to start with and used the ordinary Poisson bracket on the $(2N-2)$-dimensional phase space   
    $(x^1,\ldots,x^{N-1},p_1,\ldots,p_{N-1})$, the answer would have been the same. This is because the constraints are particularly simple, they are linear in $x^N$ and $p_N$.
In the next section we consider a slightly more complicated example with non-linear constraints.
    
Note that in the above analysis Dirac brackets are not necessary for calculating the time evolution of any quantity in phase space since, by construction,
\[ \{\phi_I,H_T\} \approx 0 \]
so
\[ \dot F = \{ F, H_T \} \approx \{ F, H_T \}_*. \]
But it is not true in general that
\[\{F, G \} \approx \{F, G \} _* \]
for any two functions $F$ and  $G$ on phase.

    \newpage
\subsection{$\phi(x)=\frac{1}{2}(x^a x^a -R^2)$ \label{Sphere} }

The condition $x^a x^a = R^2$ constrains the dynamics to the surface of a $(N-1)$-dimensional sphere of radius $R$ embedded in $\textbf{R}^N$.
Using a Lagrange multiplier in the Lagrangian,
\[ \widetilde L = \frac m 2 \dot x^a \dot x^a - { V }(x) - \frac{\lambda}{2} (x^a x^a - R^2),\]
the equations of motion are
\[ m \ddot x^a =-\frac{\partial  V }{\partial x^a} - \lambda x^a\]
from which
\beq \lambda = -\frac 1 {R^2}  \bigl(m x^a \ddot x^a + x^a \partial_a { V }\bigr).\label{lambdaR}\eeq

For the total Hamiltonian we take 
  \beq H_T(x,p) = H(x,p)+u \phi(x)=\frac{1}{2 m} p_a p_a +{ V }(x) + u \phi(x),\label{ConstrainedH}\eeq
with $p_a = \frac{\dot x^a}{m}$, with the primary constraint $\phi(x)=\frac{1}{2}(x^a x^a -R^2)$.
The time evolution of the primary constraint is
\[ \dot \phi = x^a \frac{\partial H_T}{\partial p_a}=\frac{x^a p_a}{m}\]
giving the secondary constraint
\[ \chi = x^a p_a\approx 0. \]
Note that $n^a = \frac{x^a}{|x|}\approx \frac{x^a}{R}$ is the unit normal to the sphere, so $\chi ={\mathbf x}.{\mathbf p}\approx 0$ simply demands that $p_a$ is tangential to the sphere.
Now check
\begin{align*}\dot \chi &= p_a \frac{\partial H_T}{\partial p_a} - x^a\frac{\partial H_T}{\partial x^a}
    =\frac{p_a p_a}{m} + p_a \frac{\partial u}{\partial p_a}\phi(x)  -x^a \frac{\partial { V }}{\partial x^a} -x^a \frac{\partial u}{\partial x^a}\phi(x)  - u x^ax^a\\
&\approx \frac{{\mathbf p}.{\mathbf  p}}{m}   - ({\mathbf x}.\nabla { V })_{x^2=R^2}   - u R^2.     \end{align*}
Hence $\dot\chi\approx 0$ can be satisfied by choosing
$u\approx U$ with
\beq U=\frac{1}{R^2}\left(\frac{{\mathbf p}.{\mathbf  p}}{m}   -
      ({\mathbf x}.\nabla { V })_{x^2=R^2}\right),\label{UR}\eeq
  and there are no further constraints.   $({\mathbf x}.\nabla { V })_{x^2=R^2}$ is the derivative of ${ V }$ in the direction normal to the sphere.
  Also note that $\lambda$ in (\ref{lambdaR}) and $U$ in (\ref{UR}) are not equal, rather they are related by
  \[ \lambda =  -\frac 1 {R^2}
  \left(\frac {d} {d t} (m x^a \dot x^a) -m \dot x^a \dot x^a + x^a \partial_a { V }\right)
  = U -\frac{d}{d t} \left(\frac{{\mathbf x}.{\mathbf p}}{R^2}\right),
  \]
the function $U$ in the total Hamiltonian is not the same as the Lagrange multiplier in the Lagrangian formulation, it differs by a total derivative.

  To obtain Dirac brackets
 let $\phi_1=\phi$ and $\phi_2=\chi$ giving
  \[ \{\phi_1,\phi_2\} = x^a \frac{\partial \phi_2}{\partial p_a}=x^a x^a \approx R^2\] 
  and $C_{I J}=\{\phi_I,\phi_J\}$ is
  \[ C=x^2 \begin{pmatrix} 0 & 1 \\ -1 & 0 \end{pmatrix}, \qquad
  C^{-1}=\frac{1}{x^2} \begin{pmatrix} 0 & -1 \\ 1 & 0 \end{pmatrix}.\]
Again the Dirac bracket is defined as in (\ref{DiracBracket}) giving
\begin{align*} \{F, G\}_* &  = \{F, G\}\\
  & \hskip -10pt -
    \frac{1}{x^2}
      \Bigl(-x^a\frac{\partial F}{\partial p_a}, 
    -p_a\frac{\partial F}{\partial p_a}+x^a\frac{\partial F}{\partial x^a}
    \Bigr)
\begin{pmatrix} 0&-1 \\ 1&0 \end{pmatrix}
  \begin{pmatrix}
     x^a\frac{\partial G}{\partial p_a}\\
      p_a\frac{\partial G}{\partial p_a}-x^a\frac{\partial G}{\partial x^a}
    \end{pmatrix}.\end{align*}
    In particular
    \begin{align}
      \{ x^a,x^b\}_*&= \{x^a,x^b\}=0\\
      \{ x^a,p_b\}_*&= \delta^a{}_b-\frac{x^a x^b}{x^2} \approx  \delta^a{}_b-\frac{x^a x^b}{R^2} \label{SphereMetric} \\
      \{ p_a,p_b\}_*&=\frac{1}{x^2}\bigl(x^a p_b-p_a x^b\bigr) \approx
                         \frac{1}{R^2}\bigl(x^a p_b-p_a x^b\bigr). 
      \end{align}
      This last condition is related to angular momentum, \textit{e.g.} for $N=3$
      \[ J^a = \epsilon^{a b c} x^b p_c \qquad \mbox{and} \qquad \{ p_a,p_b\}_* \approx \frac{\epsilon^{a b c} J_c}{R^2} .\] 

      \subsection{A general constraint $\phi(x)= 0$ on configuration space}

      It is straightforward to develop the ideas of \S\ref{Sphere} to a more general constraint $\phi(x)=0$ on $N$-dimensional Euclidean with Cartesian co-ordinates.
      With $\phi_1 = \phi$ one finds
      \[ \dot \phi \approx \frac{p_a \partial_a \phi}{m}\]
      so
      \[ \chi= p_a \partial_a \phi \approx 0\]
      and \[ \dot \chi \approx \frac 1 m p_a p_b \partial_a \partial_b \phi - \partial_a \phi \partial_a { V } - u \partial_a \phi \partial_a \phi \]
      so
      \[ u \approx U =  \frac{1}{|\nabla \phi|^2}\left(\frac 1 m p_a p_b \partial_a \partial_b \phi - \partial_a \phi \partial_a { V }\right).\]
      With
      \[ \{\phi,\chi \}\approx \nabla \phi .\nabla \phi\]
and       \[ C = \nabla \phi .\nabla \phi \begin{pmatrix} 0 & 1 \\ -1 & 0 \end{pmatrix}, \qquad
        C^{-1} = \frac{1}{\nabla \phi .\nabla \phi} \begin{pmatrix} 0 & -1 \\ 1 & 0 \end{pmatrix}.\]
      The Dirac bracket on phase space co-ordinates evaluates to
      \begin{align}
      \{ x^a,x^b\}_*&= \{x^a,x^b\}=0,\\
        \{ x^a,p_b\}_*&= \delta^a{}_b -  \frac{\partial_a \phi \partial_b \phi} {|\nabla \phi|^2},
                        \label{GeneralMetric} \\
        \{ p_a,p_b\}_*&=\frac{1}{|\nabla \phi|^2} p_c
                        \bigl(\partial_a  \phi \partial_b \partial_c \phi - \partial_b  \phi \partial_a \partial_c \phi \bigr).\label{Generalpp}
      \end{align}
                      
      \section{Constraints on phase space}
      
      In the previous example in \S\ref{Plane} the Lagrange multiplier $\lambda$ did not appear in the Hamiltonian, yet it is treated as a variable in the Lagrangian. This was because the canonical momentum associated to $\lambda$,
      $p_\lambda = \frac{\partial L}{\partial \dot \lambda}$ vanished, so we just ignored it.
      We could try to include $p_\lambda$ in phase space and impose $p_\lambda=0$ as further primary constraint in a phase space with $(2N+2)$ dimensions, but this presents a problem because $p_\lambda$ can not then be inverted to find $\dot \lambda$ as a function on phase space, and we need to know $\dot \lambda$ as a function on phase space in order to derive a Hamiltonian from the Lagrangian.
      
        This is an example of a general phenomenon that there can be constraints on phase space, even without a holonomic constraint on configuration space, 
        if the relation between the $p_a$ and the $\dot q^a$is not invertible, \textit{i.e.} if the functions $p_a(q,\dot q)$
        cannot be inverted to get $\dot q^a(q,p)$ in order to express the Hamiltonian
        \[ H(q,p)= p_a \dot q^a - L(q,\dot q)\]
        as a function of $q^a$ and $p_a$.  Generically this happens if the matrix $\frac{\partial^2 L}{\partial \dot q^a\partial \dot q^b}$ has a zero eigenvalue and is not invertible.
        Indeed this is really the most interesting case as holonomic constraints on configuration space do not present any real difficulty, the Lagrangian formalism is well designed for handling such holonomic constraints.
        More general constraints on phase space, involving the $p$'s, are really the focus of these notes --- \S\ref{Plane} and \S\ref{Sphere} were just warm-ups. So now we shall examine an example where primary constraints are generated by the form of the Lagrangian and are not simply imposed externally as
        holonomic constraints on configuration space.

        \subsection{Particle in a strong magnetic field\label{Strong-B}}
        
      This example is that of a charged particle moving in a 2-dimensional plane in the presence of a very strong uniform magnetic field normal to the plane.

        The Lagrangian is
        \[L = \frac m 2   \dot {\mathbf x} .\dot {\mathbf x}  + e \mathbf A(\mathbf x) . \dot {\mathbf x} -{ V }(\mathbf x)
        \]
with $\mathbf x = (x^1,x^2)$, where $e$ is the electric charge and  $\mathbf A(\mathbf x)$ the vector potential for the constant magnetic field $B$.
We take
\[ A_1 = -\frac B 2  x^2, \quad  A_2 = \frac B 2  x^1\qquad \Rightarrow \qquad B=\frac{\partial A_2}{\partial x^1}-
\frac{\partial A_1}{\partial x^2}\]
as our vector potential generating a uniform and constant magnetic field B normal to the plane. 

If $B$ is very large the $e \mathbf A(\mathbf x)$ term in the Lagrangian will dominate over the kinetic term and
\[L = e \mathbf A(\mathbf x) . \dot {\mathbf x} -{ V }(\mathbf x)=\frac{e B}{2}(x^1 \dot x^2 - x^2 \dot x^1) -{ V }(x^1,x^2).\]
The Lagrangian equations of motion are linear in the velocities,
\begin{align}
  \frac{d}{d t} \left(-\frac{e B}{2} x^2\right) = \frac{e B}{2} \dot x^2-\frac{\partial { V }}{\partial x^1}
    \quad & \Rightarrow \quad \dot x^2
 = \frac{1}{e B}\frac{\partial{ V }}{\partial x^1}\label{x2dotB}\\
  \frac{d}{d t} \left(\frac{e B}{2} x^1\right) = -\frac{e B}{2} \dot x^1-\frac{\partial { V }}{\partial x^2}
    \quad & \Rightarrow \quad 
  \dot x^1 = -\frac{1}{e B}\frac{\partial { V }}{\partial x^2} \label{x1dotB}
\end{align}
             (for example if ${ V }$ is quadratic
             \[{ V }=\frac{k}{2} \bigl\{(x^1)^2 + (x^2)^2\bigr\} \]
             the particle moves in a circle centred on the origin in the plane at the cyclotron frequency
             $\omega =\frac{k}{e m}$).
             
The canonical momenta are
\begin{align} p_1 &= -\frac{e B}{2} x^2 \label{p1B}\\p_2 &= \frac{e B}{2} x^1\label{p2B}\end{align}
and these cannot be inverted to determine $\dot x^1$ and $\dot x^2$.
However the velocities drop out of the Hamiltonian
\[ H = p_1 \dot x^1 + p_2 \dot x^2 - L = { V }(x^1,x^2),\]
which is independent of $p_1$ and $p_2$.
But the resulting Hamiltonian equations of motion
\begin{align*}  \dot x^1 &=0\\
  \dot x^2 &=0\\
  \dot p_1 &= -\frac{\partial { V }}{\partial x^1}\\
\dot p_2 &= -\frac{\partial { V }}{\partial x^2},\end{align*}
are inconsistent with (\ref{p1B}) and (\ref{p2B}) if ${ V }$ is non-trivial. The na\"{\i}ve Hamiltonian dynamics is simply wrong!

This inconsistency can be removed by imposing (\ref{p1B}) and (\ref{p2B}) as primary constraints on phase space
\begin{align} \phi_1&=p_1 +\frac{e B}{2} x^2 \approx 0,\label{p1BPS}\\
  \phi_2&=p_2 - \frac{e B}{2} x^1\approx 0,\label{p2BPS}\end{align}
with Poisson bracket
\[ \{\phi_1,\phi_2\} = eB.\] 
The total Hamiltonian is then
\[ H_T = H + u_1 \phi_1 + u_2 \phi_2= { V }(x^1,x^2)
  + u_1 \phi_1 + u_2 \phi_2\]
with $u_1(q,p)$ and $u_2(q,p)$ undetermined functions.
The time evolution using $H_T$ and ordinary Poisson brackets is then
\begin{align*} \dot \phi_1 &= \{\phi_1,H\} + \{\phi_1, u_1\phi_1 + u_2 \phi_2\} \approx -\frac{\partial { V }}{\partial x^1}+ e B u_2\\
\dot \phi_2 &= \{\phi_2,H\} + \{\phi_2, u_1\phi_1 + u_2 \phi_2\} \approx -\frac{\partial { V }}{\partial x^2} - e B u_1.
\end{align*}
These do not generate new constraints, instead they determine the unknown functions $u_1$ and $u_2$ weakly,
\[ u_1 \approx U_1=-\frac{1}{e B} \frac{\partial { V }}{\partial x^2}, \qquad
  u_2 \approx U_2=\frac{1}{e B} \frac{\partial { V }}{\partial x^1}.\]

The matrix $C_{I J} =\{\phi_I,\phi_J\}$, with $I,J=1,2$ is
\[ C = e B\begin{pmatrix} 0 & 1 \\ -1 & 0\end{pmatrix}\]
and the Dirac bracket is defined to be
\[ \{F,G\}_* = \{F,G\} - \{F,\phi_J \} (C^{-1}){}^{I J} \{ \phi_J,G\}.\]
For any functions $F$ and $G$ on phase space
\[ \{\phi_I,F\}=-\frac{\partial F}{\partial x^I} + \frac{1}{2} e B \epsilon_{I J}\frac{\partial F}{\partial p_J}\qquad \mbox{with}\qquad \epsilon_{IJ}=\begin{pmatrix}0 & 1\\ -1 & 0  \end{pmatrix}\]
so the Dirac bracket is
\begin{align*}\{ F, G \}_* &= \{ F, G\} - \frac{1}{e B} \begin{pmatrix} -\frac{\partial F}{\partial x^1} + \frac{e B}{2}\frac{\partial F}{\partial p^2},
    & -\frac{\partial F}{\partial x^2} - \frac{e B}{2}\frac{\partial F}{\partial p^1}\end{pmatrix}
  \begin{pmatrix}0 & -1\\ 1 & 0  \end{pmatrix}
  \begin{pmatrix} \frac{\partial G}{\partial x^1} - \frac{e B}{2}\frac{\partial G}{\partial p_2}\\
    \frac{\partial G}{\partial x^2} + \frac{e B}{2}\frac{\partial G}{\partial p_1}\end{pmatrix}\\
                           &=\{F, G\} - \frac{1}{e B}\left(\frac{\partial F}{\partial x^1}\frac{\partial G}{\partial x^2}
                             -\frac{\partial F}{\partial x^2}\frac{\partial G}{\partial x^1}\right)
                             - \frac{e B}{4} \left(\frac{\partial F}{\partial p_1}\frac{\partial G}{\partial p_2}
                             - \frac{\partial F}{\partial p_2}\frac{\partial G}{\partial p_1}\right)\\
                           & \hskip 120pt - \frac{1}{2} \left(\frac{\partial F}{\partial x^1}\frac{\partial G}{\partial p_1}+\frac{\partial F}{\partial x^2}\frac{\partial G}{\partial p_2} - \frac{\partial F}{\partial p_1}\frac{\partial G}{\partial x^1}-\frac{\partial F}{\partial p_2}\frac{\partial G}{\partial x^2}\right) .\end{align*}            
In particular
\begin{align*}
  \{ x^1,x^2\}_* & =-\frac{1}{e B},\\
  \{ x^1,p_1\}_* & = \{ x^2,p_2\}_* =\frac 1 2,\\
    \{ x^1,p_2\}_* & = \{ x^2,p_1\}_* = 0,\\
  \{ p_1,p_2\}_* & =-\frac{e B}{4}.\\
\end{align*}

The total Hamiltonian can be written as
\[ H_T = U_1 \phi_1 + U_2 \phi_2 + { V }(x^1,x^2)=\frac{1}{e B} \bigl(-\phi_1 \partial_2 { V } + \phi_2\partial_1 { V } \bigr)+{ V }(x^1,x^2)\]
and the Hamiltonian equations of motion are 
\begin{align*}
  \dot x^I &= \{ x^I, H_T \}_* \approx -\frac{\epsilon_{I J}}{eB} \partial_J { V }\\
  \dot p_I &  = \{ p_I, H_T \}_* \approx -\frac{1}{2} \frac{\partial { V }}{\partial x^I},
\end{align*} 
which are perfectly consistent with (\ref{x1dotB}), (\ref{x2dotB}), (\ref{p1B}) and (\ref{p2B}).

Observe again that Dirac brackets are not necessary for calculating the time evolution of any quantity in phase space since, by construction,
\[ \{\phi_I,H_T\} \approx 0 \quad \Rightarrow \quad  \dot F = \{ F, H_T \} \approx \{ F, H_T \}_*\approx
  \{F, H\}_*,\]
but $\{F,H\}$ should not be used.

\section{Dirac brackets}
      What is happening here is that Poisson brackets represent a geometrical structure on $2N$-dimensional phase space, much as a metric is a geometric structure on $N$-dimensional configuration space.  A line element on configuration space
      \[ d s^2 = g_{a b}(q) d q^a d q^b\]
      requires a metric, a symmetric $N \times N$
      matrix with components $g_{a b}$ in the chosen co-ordinate system. On phase space we can combine the $p$'s and $q$'s into a single row vector with a unified notation
      \[ X^A =(q^1,p_1,q^2,p_2,\ldots ,q^N,p_N), \]
      with $A=1,\ldots ,2N$.
      Then the Poisson bracket is
      \[ \{ X^A, X^B \} = \Omega^{A B}\]
      with the anti-symmetric $2N \times 2N$ matrix 
      \beq \Omega = \begin{pmatrix}
          0 & 1 & 0 & 0 & \cdots & 0 & 0 \\
          -1 &0 & 0 & 0 & \cdots & 0 & 0 \\
          0 & 0 & 0 & 1 & \cdots & 0 & 0 & \\
          0 & 0 & -1 & 0 & \cdots & 0 & 0 & \\
          \vdots & \vdots & \vdots & \vdots & \ddots  & \vdots & \vdots \\
          0 & 0 & 0 & 0 & \cdots & 0 & 1  \\
          0 & 0 & 0 & 0 & \cdots & -1 & 0 
        \end{pmatrix}.\label{Darboux}\eeq
        In this notation the Poisson bracket of any two functions $F(X)$ and $G(X)$ on phase space is
        \[ \{ F,G\} = \Omega^{A B}\partial_A F \partial_B G.\]
                $\Omega$ represents a geometrical structure on phase space, called a \textit{symplectic structure} and phase space is sometimes called a \textit{symplectic space}.\footnote{Mathematicians define an anti-symmetric product for infinitesimals,
          \hbox{$d X^A \wedge d X^B = - d X^B \wedge d X^A$}, called a \textit{wedge product} and define a \textit{symplectic form}, $\Omega = \frac 1 2 \Omega_{A B} d X^A \wedge d X^B$, on phase space.} There is a mathematical theorem (Darboux's theorem --- which we shall not prove here) that it is always possible to find co-ordinates on phase space so that $\Omega$ has the simple form (\ref{Darboux}) on any co-ordinate patch in phase space.\footnote{For a metric in Euclidean space it is always possible to find co-ordinates so that $g_{a b}$ is the identity matrix $\delta_{a b}$ at a single point, but this cannot be done in an extended region unless the space is flat.}
        Co-ordinates in which $\Omega$ has this form are called \textit{canonical co-ordinates} and transformations on phase space which preserve this form are called \textit{canonical transformations}. It is natural to use canonical co-ordinates on phase space whenever possible, but sometimes it is convenient to use non-canonical co-ordinates, in which case the entries in $\Omega$ become functions.

        One way of obtaining a metric on a $(N-1)$-dimensional hypersurface in Euclidean space ${\mathbf R}^N$ with Cartesian co-ordinates $x^a$ is to impose a constraint $\phi(x)=0$ on the co-ordinates. Then $\nabla \phi$ is normal to the hypersurface and
        \[ n_a = \frac{\partial_a \phi}{\sqrt{\nabla \phi . \nabla \phi}}\]
          is a unit normal.
          A metric $g_*$ on the hypersurface can be obtained by projecting the
          Euclidean metric $g_{a b}= \delta_{a b}$ in ${\mathbf R}^N$ onto
          the hypersurface
          \beq g_{*,a b} = P_a{}^{a'} P_b{}^{b'} \delta_{a' b'}\label{eq:g*}\eeq
          with the matrix
          \[P_a{}^{a'} = \delta_a{}^{a'} -  n_a n^{a'}.\]
          Since ${\mathbf n}.{\mathbf n}=1$, $P^2 = P$ is a projector.
          The metric on the hypersurface can be represented by
          \beq g_{*,a b} = \delta_{a b} - \frac{\partial_a \phi\, \partial_b \phi}{\nabla \phi . \nabla \phi}.\label{eq:G*-del}\eeq
                   For a sphere, for example, $\phi=\frac{1}{2}\bigl(x^a x^a - R^2\bigr)$, $n^a = \frac{x^a}{\sqrt{x^2}}$ and
          \[ g_{*,a b} = \delta_{a b} - \frac{x^a x^b}{R^2}\]
          which is the Dirac bracket $\{x^a,p_b \}_*$ in the example in \S\ref{Sphere}
          (Note that $ g_{*,a b}$ is still an $N\times N$ matrix in this formalism, but it has a zero eigenvalue and is of rank $N-1$).
          For the more general case (\ref{eq:G*-del}) is the Dirac bracket (\ref{GeneralMetric}),
          while (\ref{Generalpp}) can be written as
          \[\{ p_a , p_b \}_* \approx n_a \partial_b (\mathbf{p}.\mathbf{n}) - n_b \partial_a (\mathbf{p}.\mathbf{n}).\]

          For a constrained system in phase space with ${\cal J}$ constraints the Dirac bracket represents a projection of the symplectic structure $\Omega$ in the $2N \times 2N$ dimensional phase space onto a symplectic structure $\Omega_*$ on the $(2N-{\cal J})\times (2N-{\cal J})$ symplectic space of the constrained system (for Dirac brackets to be defined ${\cal J}$ must be even). $\Omega_*$ is a $2N\times 2N$ matrix but it has ${\cal J}$ zero eigenvalues and is of rank $(2N-{\cal J})\times (2N-{\cal J})$.
          For the linear example in \S\ref{Plane} the canonical co-ordinates $(x^a,p_b)$ for $\Omega$ project onto canonical co-ordinates for $\Omega_*$, but for the spherical example in \S\ref{Sphere}  $(x^a,p_b)$ do not project onto canonical co-ordinates on $\Omega_*$ and as a result $\Omega_*$ is not a constant matrix, it's entries depend on the point in phase space.\footnote{In fact, because of the high degree of symmetry of a sphere, it is possible to find canonical co-ordinates for $\Omega$ that project to canonical co-ordinates for $\Omega_*$ in \S \ref{Sphere}. Choose spherical polar co-ordinates $(r,\theta^1,\ldots, \theta^{N-1})$ for ${\mathbf R}^N$ and impose the constraint $\phi=\frac{1}{2}(r^2-R^2)$.  It is left as an exercise to calculate the resulting Dirac brackets and show that $\Omega_*$ has canonical form in these co-ordinates.}

          To see how this works consider the Dirac bracket as defined in (\ref{DiracBracket})
          \[ \{ F, G\}_* = \{F, G\} - \{F,\phi_I\} \bigl(C^{-1}\bigr)^{I J} \{ \phi_J, G\},\]
          with $C^{-1}$ the inverse of the ${\cal J} \times {\cal J}$ matrix $C$ that has components
         \[C_{I J}=\{\phi_I, \phi_J \} =  \Omega^{A B}\partial_A \phi_I  \partial_B \phi_J.\]
          Define
          \[ Q_{A B} = \partial_A \phi_I (C^{-1})^{I J} \partial_B \phi_J\]
          (note that as a $2N\times 2N$ matrix $Q^T=-Q$ is anti-symmetric, since $C_{I J}$ is anti-symmetric).
          Then the Dirac bracket of two functions $F$ and $G$ on the full phase space can be written as
          \[ \{F,G\}_* = \Omega_*^{A B} \partial_A F \partial_B G\]
          with
          \[\Omega_*^{A B} = \Omega^{A B} -\Omega^{A C} Q_{C D} \Omega^{D B}\]
          or \beq\Omega_* = \Omega - \Omega Q \Omega\label{eq:Omega*}\eeq
          as a matrix equation.
          Now observe that
          \begin{align*}
            (Q \Omega Q)_{A B}
         &= \partial_A \phi_I (C^{-1})^{I J} \partial_C \phi_J \Omega^{C D} \partial_D \phi_K (C^{-1})^{K L} \partial_B \phi_L\\
         &=\partial_A \phi_I (C^{-1})^{I J} C_{J K} (C^{-1})^{K L}\partial_B \phi_L \\
            & =\partial_A \phi_I (C^{-1})^{I L}\partial_B \phi_L\\
            & = Q_{A B}.
                                 \end{align*}
     Thus $Q\Omega Q = Q$, or
     \[ (Q \Omega)^2 = Q\Omega\]
     and $Q \Omega$ is a projection operator.
  Similarly
     \[ (\Omega Q)^2=\Omega Q\]
  and, since $Q$ and $\Omega$ are both anti-symmetric,
  \[ (\Omega Q)^T =Q^T \Omega^T  = Q \Omega.\] 
 Equation (\ref{eq:Omega*}) can now be written as                            
 \[ \Omega_* = (1-\Omega Q)\Omega (1- Q \Omega)\]

     Since $\Omega Q$ is a projection operator so is  ${\cal P} = 1-\Omega Q$, ${\cal P}^2={\cal P}$, (but ${\cal P}$ is not symmetric unless $Q$ commutes with $\Omega$, which need not be true).
     In any case
     \[ \Omega_* = {\cal P} \Omega {\cal P}^T\]
     or, in co-ordinates,
     \[ \Omega^{A B}_* = {\cal P}^A{}_C {\cal P}^B{}_D \Omega^{C D},\]
 an identical structure to (\ref{eq:g*}).
     
            \section{More general constraints on phase space}
            
             Suppose there is a number of constraints on phase space, $\phi_m(q,p)=0$ with $m=1,\ldots,{\mathfrak M}$ (perhaps some holonomic constraints on configuration space but definitely some constraints that mean that not all the $p_a$ can be inverted to get $\dot q^a$ as functions on phase space).  In the Lagrangian formulation consider the extended Lagrangian
        \[ \widetilde L(q,\dot q,p) = L(q,\dot q)- u_m(q,\dot q)  \phi_m(q,p)=  p_a \dot q^a - H(q,p)  - u_m(q,\dot q)  \phi_m(q,p) \]
   where $u_m$ are undetermined functions of $q^a$ and $\dot q^a$.
Now apply a variational principle to $\widetilde L$ that extremises $\widetilde S=\int \widetilde L dt$,
     \begin{align*}
         \delta \widetilde L &= p_a  \delta \dot q^a
          + \dot q^a \delta p_a 
          -\left(\frac{\partial H}{\partial q^a} + u_m\frac{\partial \phi_m}{\partial q^a}\right)\delta q^a
          -\left(\frac{\partial H}{\partial p_a} +u_m\frac{\partial \phi_m}{\partial p_a}\right)\delta p_a
          - \phi_m \delta u_m \\
                                       &\approx p_a \delta \dot q^a
          - \left(
                    \frac{\partial H}{\partial q^a}  u_m\frac{\partial \phi_m}{\partial q^a}\right)\delta q^a
           +\left(\dot q^a -\frac{\partial H}{\partial p_a} -u_m\frac{\partial \phi_m}{\partial p_a}\right)\delta p_a\\
                          & =\frac {d (p_a  \delta q^a )} {d t} 
                            - \left(\dot p_a 
 +\frac{\partial H}{\partial q^a} + u_m\frac{\partial \phi_m}{\partial q^a}\right)\delta q^a
            +\left(\dot q^a -\frac{\partial H}{\partial p_a} -u_m\frac{\partial \phi_m}{\partial p_a}\right)\delta p_a.
       \end{align*}
With $\delta q^a$ fixed at the end points $\widetilde S$ is extremised when
\[\dot p_a=
  -\frac{\partial H}{\partial q^a} - u_m\frac{\partial \phi_m}{\partial q^a}, \qquad
\dot q^a =\frac{\partial H}{\partial p_a} + u_m\frac{\partial \phi_m}{\partial p_a}.\]
      We also have that
      \[ \frac{\delta \widetilde L}{\partial \dot q^a}= \frac{\delta L}{\partial \dot q^a}- \frac{\delta  u_m}{\partial \dot q^a}\phi \approx \frac{\delta L}{\partial \dot q^a}, \]
      so we can use either
      \[ p_a = \frac{\widetilde \delta L}{\partial \dot q^a} \quad \mbox{or}\quad  p_a=\frac{\delta L}{\partial \dot q^a}.\]
      This analysis does not require the explicit form of
      \[\delta u_m= \left(\frac{\partial u_m}{\partial \dot q^a} \delta \dot q^a
          + \frac{\partial u_m}{\partial q^a} \delta q^a\right) \] 
      but it is essential that all variations are made \textit{before} applying the constraints.
      
      The upshot is that the equations of motion in Hamiltonian form can be written as
      \begin{align}\dot p_a&=
  -\frac{\partial H}{\partial q^a} - u_m\frac{\partial \phi_m}{\partial q^a}, \\
\dot q^a &=\frac{\partial H}{\partial p_a} + u_m\frac{\partial \phi_m}{\partial p_a}.\end{align}

In terms of the total Hamiltonian
\[H_T=H+u_m \phi_m\]
the time evolution of any function $F(q,p)$ on phase space
can be written as
\[ \dot F = \{ F, H_T\} = \{F,H\} + \{F,  u_m\}\phi + u_m\{ F , \phi_m\}\approx
  \{ F, H \} + u_m\{ F , \phi_m\}.\]
This is a little sloppy as the Poisson bracket $\{ F, u_m\}$ does not exist if $u_m$ depends on any velocity $\dot q^a$ that cannot be expressed as a function on phase space, in which case
$ \{ F, H_T \} $ simply does not exist off the constraint surface,
but we shall assume that
\[ \dot F \approx
  \{ F, H \} + u_m\{ F , \phi_m\}\]
is still true even then.

\subsection{First class constraints \label{1stClass}}

            In \S\ref{Strong-B} we saw an example with two constraints $\phi_I(q,p)$ on phase space which did not come from an externally imposed constraint on configuration space but instead 
were dictated by the form of $L$ itself: $p^a(q,\dot q)$ could not be inverted to give $\dot q^a(q,p)$. 
                       There is another kind of constraint on phase space that is very important with the same origin, and that is a constraint associated with a redundancy in the physical description --- what is called a \textit{gauge symmetry}.

            To illustrate this consider the following example: take an odd dimensional configuration space, $N=2n+1$ with
            \begin{align*}
              q^a(t) &= x^a(t),\qquad a = 1,\ldots , 2n\\
              q^{2N+1}(t) & = A(t),
              \end{align*}
              where $x^a$ are Cartesian co-ordinates on ${\mathbf R}^{2n}$.
              Define complex co-ordinates
              \[ z^1 = \frac{x^1 + i x^{1+n}}{\sqrt 2}, \ldots, z^n = \frac{x^n + i x^{2n}}{\sqrt 2} \]
              giving this part of configuration space the complex structure of ${\mathbf C}^n$.
              The dynamics will be defined by a Lagrangian
              \beq L = \bigl \{ (\dot z^i)^* + i A (z^i)^*\bigr\}  \bigl ( \dot z^i - i A z^i\bigr)
                -{ V }(z^\dagger z),\label{LAz}\eeq
              where now $i=1,\ldots , n$ only.

              If we define the \textit{co-variant derivative}  of $z^i$ by
              \[ D z^i = \dot z^i - i A z^i\]
              this can be re-expressed as
              \[ L=\bigl(D z \bigr)^\dagger D z - { V }(z^\dagger z).\]
           
            The dynamics is invariant under the transformation
            \[ z^i \rightarrow e^{i\theta} z^i, \qquad A \rightarrow A+\dot \theta\]
            where $\theta(t)$ is any differentiable function. Clearly this leaves $z^\dagger z$ unchanged while
            \[ D z \ \rightarrow \  e^{i\theta} \dot z^i + i \dot \theta   e^{i\theta} z^i - i (A+\dot \theta)  e^{i\theta} z^i =  e^{i\theta}(\dot z - i A z^i)= e^{i\theta} D z^i,\]
            hence $\bigl(Dz\bigr)^\dagger Dz$ is also invariant and the complete Lagrangian is invariant.

            There are no constraints being applied to configuration space here, all $2n+1$ co-ordinates can be freely varied without constraint.   The Lagrangian equations of motion are
            \[\frac{d}{d t}\left(\frac{\partial L}{\partial \dot z^i}\right)
              = \frac{\partial { V }}{\partial z^i}\quad \Rightarrow \quad \bigl(D z^i\bigr)^* = \frac{\partial { V }}{\partial z^i},\]
            plus its complex conjugate
             \[\frac{d}{d t}\left(\frac{\partial L}{\partial (\dot z^i)^*}\right)
               = \frac{\partial { V }}{(\partial z^i)^*}\quad \Rightarrow \quad D z^i = \frac{\partial { V }}{\partial (z^i)^*},\]
             and
            \[\frac{d}{d t}\left(\frac{\partial L}{\partial \dot A}\right)
              = \frac{\partial L}{\partial A}\quad \Rightarrow \quad
              0=i \bigl\{ (z^i)^*\dot z^i - z^i (\dot z^i)^*\bigr\} + 2 A z^\dagger z,\]
            which determines $A$ provided the $z^i$ do not all vanish
            \[ A = \frac{i\bigl\{  z^i (\dot z^i)^* - (z^i)^*\dot z^i\bigr\}}{2 z^\dagger z}.\]
            This is not a constraint, it is an equation of motion: $A$ is not a Lagrange multiplier because $L$ is not linear in $A$.
            
            Nevertheless the form of (\ref{LAz}) imposes a constraint on phase space.
            The momenta conjugate to $z^i$ are\footnote{We use $\pi_i^*$ here to distinguish the complex canonical momenta conjugate to $z^i$ from $p_a$, the real canonical momenta conjugate to $x^a$. The $*$ on $\pi_i^*$ is because, in the absence of $A$, $L=\frac{1}{2}\dot x^a \dot x^a -{ V }$ and $p_a = \dot x^a$ with
              $\{ x^a , p_b\}=\delta^a{}_b$. Hence, for example, $\{ \frac{x^1 + i x^2}{\sqrt 2}, \frac{p_1 - i p_2}{\sqrt 2 }\}=1$. In complex co-ordinates this translates to $\{z^1, \pi_1^* \} =1$ in complex variables, hence $\pi_1^*$ is conjugate to $z^1$. In general  $\{z^i, \pi_j^* \} =\delta^i{}_j$}
            \[ \pi_i^* = \frac{\partial L}{\partial \dot z^i} = \bigl( D z^i \bigr)^*\] 
            and, since $z^i$ and $(z^i)^*$ should be considered independent, there is also
             \[ \pi_i = \frac{\partial L}{\partial (\dot z^i)^*} = D z^i. \] 
            Denote the momentum conjugate to $A$ by $\Pi$, then\footnote{We don't use $p_A$ in case $A$ gets confused with an index.} 
            \[ \Pi = \frac{\partial L}{\partial \dot A}=0\]
            giving a primary constraint in phase space, $\phi \approx 0$ with $\phi = \Pi$.
            Since $\Pi(A,\dot A)=0$ we cannot invert it to get $\dot A(A,\Pi)$.
                   
            $\dot A$ is arbitrary, it is not determined by the Lagrangian,
            but since $\Pi=0$, we can write the na{\"\i}ve Hamiltonian
            \[ H = \pi^\dagger \pi +{ V }(z^\dagger z)+  i A(\pi^\dagger z - z^\dagger \pi) \]
            so the total Hamilton is
            \beq H_T = H +  u \phi,\label{Gauged-Hamiltonian}\eeq
            with $u$ an arbitrary function, and impose the constraint $\phi=\Pi\approx 0$.
Again we must require that the time derivative of $\phi$ is consistent
\[ \dot \phi = \{ \phi,H_T\} = -\frac{\partial H_T}{\partial A} \approx  -i (\pi^\dagger z - z^\dagger \pi),\]
giving a secondary constraint
\[ \chi \approx 0 \quad \mbox{with} \quad \chi = i(z^\dagger \pi- \pi^\dagger z).\]
Continuing
\begin{align*} \dot\chi = \{ \chi,H_T\}
  &= i\left\{ -\pi_i^* \frac{\partial H_T}{\partial \pi_i^*} 
                         + z^i \frac{\partial H_T}{\partial z^i}
                          +  \pi_i \frac{\partial H_T}{\partial \pi_i}
                          -  (z^i)^* \frac{\partial H_T}{\partial (z^i)^*}
    \right\}\\
  & \approx  i\left\{ \left(-\pi^\dagger \pi - i A \pi^\dagger z\right) +
    \left(z^i \frac{\partial { V }}{\partial z^i} 
    +i A \pi^\dagger z\right)\right.\\
  & \hskip 30pt
\left.    +\left(\pi^\dagger \pi  - i A z^\dagger \pi\right)
                          -\left((z^i)^* \frac{\partial { V }}{\partial (z^i)^*} -iA z^\dagger \pi\right)
                          \right\} + u\{ \chi,\phi\}\\
  &=-   z^i \frac{\partial { V }}{\partial z^i} +  (z^i)^* \frac{\partial { V }}{\partial (z^i)^*}  + u\{ \chi,\phi\}.
\end{align*}
Now, with $Z=z^\dagger z$,
\[ z^i \frac{\partial { V }}{\partial z^i} -  (z^i)^* \frac{\partial { V }}{\partial (z^i)^*}
    = z^i (z^i)^* \frac{d { V }}{d Z} - (z^i)^* z^i  \frac{d { V }}{d Z}=0\]
  and hence
  \[ \dot \chi \approx  u\{ \chi,\phi\}. \]
  Obviously setting $u\approx 0$ gives a consistent time evolution with no new constraints, but this is not the only possibility. Since the Poisson bracket
  \[\{\chi,\phi\}=0\]
$u$ can actually be any function, consistent time evolution does not constrain $u$ as in the previous examples, $u$ remains arbitrary.

  We are left with two constraints,
  \beq \phi = \Pi\qquad \mbox{and} \qquad  \chi = i(z^\dagger \pi- \pi^\dagger z),\label{phi-chi-CPn}\eeq
  which commute 
  \[ \{\phi,\chi\}=0,\]
so the matrix $C=0$ vanishes strongly and the technique of using Dirac brackets cannot be applied straightforwardly.
 
What is happening here is that there is a redundant degree freedom in the description of the dynamics in terms of $z^i$ and $A$ that is not physical, we can change $z^i \rightarrow e^{i\theta} z^i$ and $A\rightarrow A + \dot \theta$ for any arbitrary $\theta(t)$ and the physics is unchanged. This means that any description of the dynamics must contain an arbitrary function that contains no physical information, it also means that the matrix $C$ is not invertible.  This is in contrast to the constraints studied in \S\ref{Plane}, \S\ref{Sphere} and \S\ref{Strong-B}
where everything was determined and $C$ was invertible. The distinction between these two cases is important and a constraint that commutes with all the other constraints, using Poisson brackets,
is called \textit{first class} while a constraint that does not commute with all the other constraints is called \textit{second class}.  All the constraints in \S\ref{Plane}, \S\ref{Sphere} and \S\ref{Strong-B} were second class while $\phi_1$ and $\phi_2$ in (\ref{Extened-Hamiltonian}) are first class.

We can try to remove this redundancy by fixing $\theta$. For example if $z^n\ne0$ we could choose $\theta$ to make $z^n$ real and positive, which is the same thing as setting $x^{2n}=0$. This can be achieved by adding the configuration space constraint
\[ \eta= i(z^n - (z^n)^*)\approx 0\]
to the Hamiltonian as another primary constraint and defining a ``gauge fixed'' Hamiltonian
\beq H_\eta= H+ u_1 \phi  + u_2 \eta,\label{CPnHamiltonian'}\eeq
where $u$ has been re-labelled $u_1$ and a second unknown function $u_2$ has been introduced.
We should check the time evolution of the two primary constraints $\phi$ and $\eta$ using the gauge fixed Hamiltonian,
\[ \dot \phi = \{\phi, H_\eta\} \approx -i \bigl(\pi^\dagger z - z^\dagger \pi\bigr) + u_2\{\phi,\eta\}= \chi + u_2\{\phi,\eta\}.\]
Since $\{\phi,\eta\}=0$
\[\dot \phi \approx 0 \quad \Rightarrow \quad \chi \approx 0\] as before, with $u_2$ arbitrary.

Using the gauge fixed Hamiltonian the time evolution of $\chi$ is
\[ \dot \chi = \{ \chi, H_\eta \}=\{ \chi, H\} + u_1 \{ \chi,\phi\} + u_2 \{ \chi,\eta\} \approx u_1\{\chi,\phi\}-u_2\bigl((z^n + (z^n)^*\bigr),\]
since $\{\chi,H\}=0$.  But $\{\chi,\phi\}=0$ as before so now 
and consistent time evolution requires that
\[ \dot\chi \approx 0 \quad \Rightarrow \quad u_2 \approx 0\]
with $u_1$ arbitrary.
Lastly
\begin{align*}
  \dot \eta &= \{ \eta, H_\eta\}= i
    \left(\frac{\partial H_\eta}{\partial \pi_n^*} -\frac{\partial H_\eta}{\partial \pi_n} \right) \\
            &\approx i\bigl((\pi_n + i A z^n - \pi_n^* + i A (z^n)^*\bigr)+ u_1 \{\eta,\phi\}, \end{align*}
          with $\{\eta,\phi\}=0$,
          leading to a fourth constraint
          \[ \delta = i\bigl( \pi_n - \pi_n^*\bigr) - A \bigl( z^n + (z^n)^* \bigr) \approx 0,\]
          with $u_1$ still arbitrary.
          \textit{i.e.}
          \beq A \approx  i\left(\frac{ \pi_n - \pi_n^*}{z^n + (z^n)^*}\right)\label{eq:A} \eeq
           which is a natural consequence of
          \[ \pi_n = \dot z^n - i A z^n,\]
          if $z^n$ is real at all times then $\dot z^n$ is real and the imaginary part of $\pi_n$ is \hbox{$-A z^n=-\frac A 2 \bigl(z^n + ( z^n)^*\bigr)$}.
          
          Finally, using $H_\eta$,
          \begin{align*}
            \dot \delta &= \{ \delta,H_\eta\} = i\left(-\frac{\partial H_\eta}{\partial (z^n)^*}
                          +\frac{\partial H_\eta}{\partial z^n}\right)- \bigl( z^n + (z^n)^* \bigr)\frac{\partial H_\eta}{\partial \Pi}-A\left(\frac{\partial H_\eta}{\partial \pi_n^*} +\frac{\partial H_\eta}{\partial \pi_n} \right)\\
                        &\approx  i \left(-\frac{\partial { V }}{\partial (z^n)^*} + i A \pi_n +\frac{\partial { V }}{\partial z^n} + i A \pi_n^* \right) +  \bigl( z^n + (z^n)^* \bigr)u_1\\
& \hskip 150pt   -A \Bigl(\pi_n + i A z_n + \pi_n^* - i A (z_n)^* \Bigr)\\
                        &                          = - i\bigl((z^n -(z^n)^*\bigr)\frac{\partial { V }}{\partial Z}
                          + \bigl( z^n + (z^n)^* \bigr) u_1 - 2 A \bigl( \pi_n + \pi_n^*\bigr)
                          - i A^2\bigl(z^n -(z^n)^*)\\
                        & \approx  \bigl( z^n + (z^n)^* \bigr) u_1 - 2 A \bigl( \pi_n + \pi_n^*\bigr)\\
            ~& \approx  \bigl( z^n + (z^n)^* \bigr) u_1 - 2 i \frac{\bigl( \pi_n + \pi_n^*\bigr)\bigl( \pi_n - \pi_n^*\bigr)}{ \bigl( z^n + (z^n)^* \bigr)},
\end{align*}
where the last line uses (\ref{eq:A}).  Defining
\[ U_1 =  2 A \left(\frac{\pi_n + \pi_n^*}{z^n + (z^n)^* }\right)\]
consistent time evolution is obtained by setting $u_1\approx U_1$
and does not produce any more constraints.

There are now four constraints:
\begin{align*}
  \phi& = \Pi,\\
  \chi&= i(z^\dagger \pi - \pi^\dagger z),\\
  \eta& = i\bigl( z^n - (z^n)^*\bigr), \\
  \delta& =  i\bigl( \pi_n - \pi_n^*\bigr) - A \bigl( z^n + (z^n)^* \bigr).
\end{align*}
Calculating all the Poisson brackets between any pair the possibilities are 
\begin{align*}
  \{\phi,\chi \}&=0, \qquad \{\phi,\eta\}=0, \qquad  \{\chi,\eta\}= - \bigl(z^n + (z^n)^*\bigr),\\
    \{\phi,\delta\}&= z^n + (z^n)^*, \qquad \{\chi,\delta\}\approx -\bigl(\pi_n + \pi_n^*\bigr), \qquad  \{\eta,\delta\}= 2.
\end{align*}

Let $\phi=\phi_1$, $\delta=\phi_2$, $\eta=\phi_3$, $\chi=\phi_4$ and $C_{I J}=\{\phi_I,\phi_J\}$,
with $I,J=1,\ldots,4$, then
\[ C = 2\begin{pmatrix} 0 & Re(z^n) & 0 & 0 \\ - Re(z^n) & 0 & -1 &  Re(\pi_n) \\ 0 & 1 & 0&  Re(z^n)\\
    0 & - Re(\pi_n) & - Re(z^n)& 0\end{pmatrix}.\]
Since
\[\det C = \{2 Re(z^n)\}^4  \] 
$C$ is invertible if $Re(z^n) \ne 0$. Provided $x^n\ne 0$ we can construct Dirac brackets on
a reduced $(2N-2)\times (2N-2)$ phase space.

While this gauge fixing procedure gives consistent Dirac brackets on a region of phase space, it is somewhat unsatisfactory as it fails when $x^n=0$ ---
which may happen for any given solution of the equations of motion as the dynamics unfolds.
As long as $z^i$ are not all zero there is always one that we can choose to set to be real and positive by a using $\theta$ to change its phase and fix the gauge appropriately, but there is no such gauge fixing that works globally in configuration space for all $(z^1,\ldots,z^n)$, in general we need to make different gauge choices in different co-ordinate patches of configuration space.

Going back to the case where the gauge is not fixed note that $\chi$ in (\ref{phi-chi-CPn}) already appears linearly in the total Hamilton $H_T$ in (\ref{Gauged-Hamiltonian}), indeed we can replace $A$ in (\ref{Gauged-Hamiltonian})
with a completely general function $\tilde u$ and define an \textit{extended Hamiltonian}
\[ H_E = \pi^\dagger \pi +{ V }(z^\dagger z)+  u  \Pi + i \tilde u (z^\dagger \pi- \pi^\dagger z),\]
For a more concise notation let $\phi=\phi_1$, $u=u_1$, $\chi = \phi_2$ and $\tilde u=u_2$ (\textit{not} the same as $u_2$ in the gauge fixed Hamiltonian above!), the extended Hamiltonian is 
 \beq H_E = \pi^\dagger \pi +{ V }(z^\dagger z)+  u_1 \phi_1+u_2 \phi_2.\label{Extened-Hamiltonian}\eeq
The extended Hamiltonian is the most general Hamiltonian, with no gauge fixing, that gives a consistent time evolution on phase space, but the evolution is not unique as $H_E$ contains two arbitrary functions.

In the particular case of the above construction with $n=1$ ($N=3$) the motion corresponds to that of a particle moving on the positive real line, with  $z^1=r  e^{i\phi}$ the phase has no physical relevance and $r> 0$.

\subsection{${\mathbf {CP}}^{n-1}$ \label{CPn}}

As an aside we note that, for the particular choice ${ V }=(z^\dagger z -R^2)^2$ the minimum energy occurs
for $z^\dagger z = \frac{R^2}{2}$, which is a $(2n-1)$-dimensional sphere $S^{2n-1}$ in ${\mathbf C}^n$.
Since all points with $z^i$ and $e^{i\theta} z^i$ are physically equivalent the minimum energy surface has the geometry of $S^{2n-1}/U(1)$, a geometry which, for $n\ge 2$, is known in the mathematical literature as the \textit{complex projective space}, ${\mathbf {CP}}^{n-1}$, an $(n-1)$-complex dimensional space ($(2n-2)$-real dimensions). For $n=2$, ${\mathbf {CP}}^1$ is isomorphic to the familiar 2-dimensional sphere $S^2$, but for $n>2$ they are more complicated complex manifolds with some very interesting properties.

We can constrain the dynamics of the whole system to ${\mathbf {CP}}^{n-1}$ by adding the holonomic constraint $\psi =z^\dagger z - \frac{R^2}{2}$ to the Hamilton (\ref{Gauged-Hamiltonian}) and using
\beq H_T = \pi^\dagger \pi + i A(\pi^\dagger z - z^\dagger \pi) + u_1\phi+  u_2 \psi,\label{CPnHamiltonian}\eeq
with $u_1$ and $u_2$ arbitrary functions
(we have set ${ V }$ to zero as it is just a constant when $z^\dagger z =R^2$).
We should check the time evolution of the primary constraints $\psi\approx 0$ and $\phi=\Pi\approx 0$.
As before
\[ \dot \phi \approx -i(\pi^\dagger z - z^\dagger \pi),\]
giving the secondary constraint
\[ \chi=i(\pi^\dagger z - z^\dagger \pi)\approx 0.\]
while
\begin{align*} \dot \psi &= \{\psi, H_T\} = z^i \frac{\partial H_T}{\partial \pi_i} + (z^i)^* \frac{\partial H_T}{\partial \pi_i^*}\\
  &=\pi^\dagger z + z^\dagger \pi  
 -i A z^\dagger z  + i A z^\dagger z \\
  &\approx \pi^\dagger z + z^\dagger \pi, \end{align*}
giving another secondary constraint
\[ \eta = \pi^\dagger z + z^\dagger \pi \approx 0.\]
Carrying on
\begin{align*}
\dot \chi & = i\left\{  \pi_i^* \left(\frac{\partial H_T}{\partial \pi_i^*}\right)
  - z^i \left(\frac{\partial H_T}{\partial z^i}\right) -   \pi_i\left(\frac{\partial H_T}{\partial \pi_i}\right)
            + (z^i)^* \left(\frac{\partial H_T}{\partial (z^i)^*}\right)\right\}\\
  & \hskip 40pt + u_1\{\chi,\phi\} +u_2\{\chi,\psi\}\\
  &\approx u_1\{\chi,\phi\} +u_2\{\chi,\psi\},\\
  \dot \eta &=
  \pi_i^* \left(\frac{\partial H_T}{\partial \pi_i^*}\right)
  - z^i \left(\frac{\partial H_T}{\partial z^i}\right) +   \pi_i\left(\frac{\partial H_T}{\partial \pi_i}\right)
              - (z^i)^* \left(\frac{\partial H_T}{\partial (z^i)^*}\right)\\
&   \hskip 40pt + u_1\{\eta,\phi\} +u_2\{\eta,\psi\}\\
  &\approx  \pi_i^* \pi_i - u z^i(z^i)^* + \pi \pi^* - u (z^i)^*z^i \approx 2\pi^\dagger \pi -  u R^2,  
\end{align*}
so
\[ u \approx U \qquad \mbox{with} \qquad U=\frac{2\pi^\dagger \pi}{R^2}.\]
and there are no further secondary constraints.

There are four constraints in all and their Poisson brackets are
\[ \{\phi,\psi\}=\{\phi,\chi\}=\{\phi,\eta\}=\{\psi, \chi \}=\{\chi, \eta \}=0, \qquad \{ \psi , \eta\}=2 z^\dagger z \approx  R^2.\]
Labelling $\psi=\phi_1$, $\eta=\phi_2$, $\phi=\phi_3$ and $\chi=\phi_4$ the $4\times 4$ matrix $C_{I J}=\{\phi_I,\phi_J\}$ is
\[ C=R^2 \begin{pmatrix} 0 & 1 & 0 & 0 \\ -1 & 0 & 0 & 0 \\ 0 & 0 & 0 & 0\\ 0 & 0 & 0 & 0\\
  \end{pmatrix} \]
which is not invertible because $\phi_3$ and $\phi_4$ are  first class.
This highlights an important distinction between first class and second class constraints, second class constraints allow the symplectic structure $\Omega$ to be projected onto a smaller dimensional  phase space, first class constraints do not.

\subsection{Relativistic particle \label{RelativisticParticle}}

Another example that generates first class constraints is an $(N+1)$-dimensional configuration space with $N$ Cartesian co-ordinates $x^i$, one extra co-ordinate $y$ and Lagrangian
\beq L=\frac 1 2 \left( \frac{\dot x^i \dot x^i}{y}  + y\right).\label{eq:RelativisticParticle}\eeq
Since $\frac{\partial L}{\partial \dot y}=0$ the equation of motion for $y$ is
\[ y^2 = \dot x^i \dot x^i.\]
This is purely algebraic and putting $y=\sqrt{\dot x^i \dot x^i}$ back into the Lagrangian reduces it to
  \[ L=\sqrt{\dot x^i \dot x^i},\]
  the action is just the length of the curve $x^i(t)$ and the minimum length
  is when $x^i(t)$ is just a straight line.

  In form (\ref{eq:RelativisticParticle}) the canonical momentum associated with $y$ is $p_y=0$.
The other momenta are
\[ p_i = \frac{\partial L}{\partial \dot x^i} = \frac{\dot x^i}{y}\]
so
\[ H = \frac{y}{2}( p^i p^i - 1).\]
Since $p_y$ vanishes it cannot be inverted to obtain $\dot y$ as a function of the co-ordinates and momenta. We add $\phi=p_y$ to $H$ as a primary constraint,
\[ H_T = \frac{y}{2}(p^i p^i -1) + u\phi\]
with $v$ an undetermined function.

The time evolution of $\phi$ is
\[ \dot \phi = -\frac{\partial H_T}{\partial y}=\frac{1}{2}(1-p^i p^i)+\dot u \phi\]
  leading to the secondary constraint
  \[ \chi = \frac{1}{2}(p^i p^i-1).\]
  Carrying on
  \[ \dot \chi = -p^i \frac{\partial H_T}{\partial x^i}\approx 0,\]
  there are no further constraints and $u$ is still undetermined.
  \[ \{ \phi,\chi\}=0\]
  and both constraints are first class.
  Note that the Hamiltonian itself
  \[ H_T = - y \chi + u \phi \approx 0\]
  vanishes weakly.

  This example is in fact very relevant to the relativistic motion of a point
  particle.  Let $N=4$ and consider Cartesian co-ordinates in 4-dimensional Minkowski space-time $x^i=(x^0,x^1,x^2,x^3)$, with metric $\eta_{i j}=
  \scriptstyle{\begin{pmatrix} 1 & 0 & 0 & 0\\ 0 & -1 & 0 & 0 \\ 0 & 0 & -1 & 0 \\ 0 & 0 & 0 & -1\end{pmatrix}}$. Parameterise a curve by $\tau$ and consider a particle following a time-like trajectory
  $x^i(\tau)$, with $\dot x^i = \frac{d x^i}{d \tau}$ and
  \[ \eta_{i j} \dot x^i \dot x^j = (\dot x^0)^2- (\dot x^1)^2- (\dot x^2)^2 - (\dot x^3)^2>0.\]
  The Lagrangian
\[ L=\frac 1 2 \left(\frac{ \eta_{i j}\dot x^i \dot x^j}{y}  + y m^2\right)\]  
is Lorentz invariant and the equation of motion for $y$ is
\[ y^2 =  \frac{\eta_{i j}\dot x^i \dot x^j}{m^2}.\]
The action is
\[ S=\frac 1 2 \int_1^2 \left(\frac{\eta_{i j}}{y^2} \frac{d x^i}{d \tau}
  \frac{d x^j}{d \tau} +m^2\right)y d\tau =\frac 1 2 \int_1^2 \left(\eta_{i j} \frac{d x^i}{d \tau'}
  \frac{d x^j}{d \tau'} +m^2\right)d\tau'\]
where $d \tau' = y(\tau) d \tau$, \textit{i.e.} $\tau' = \int y(\tau) d \tau$.
$y(\tau)$ has no physical significance, it merely reflects the freedom to parameterise the world-line however we wish.

The Lagrangian implies the constraint \[\phi=p_y\approx 0\]
and
\[ H_T=\frac{y}{2}(\mathbf p.\mathbf p -m^2) + u \phi \ \Rightarrow \ 
\dot \phi \approx  \frac{1}{2}(m^2 - \mathbf p.\mathbf p)\ \Rightarrow
\    \chi =  \frac{1}{2}(m^2 - \mathbf p.\mathbf p) \approx 0,\]
where $\mathbf p.\mathbf p=\eta^{i j} p_i p_j$ and $u$ remains arbitrary.

The fact that the Hamiltonian vanishes weakly is characteristic of
reparameterisation invariant relativistic theories (such as General Relativity) and is a consequence of the fact that defining an energy requires choosing a reference frame and no Lorentz frame has been chosen in defining the above Hamiltonian.

\section{Formal development of the Dirac formalism}

Motivated by the study of the previous examples we go on to develop the mathematical formalism of constrained dynamics in phase space.
In general there can be more than one \textit{primary} constraint, $\phi_m$, $m=0,\ldots,\mathfrak {M}$. These can be either holonomic constraints on configuration space, that do not involve the $p_a$, or they can be constraints on phase space arising from the from the form of the Lagrangian if the relations $p_a = \frac{\partial L}{\partial \dot x^a}$ are not invertible, \textit{i.e.} the $p_a(x,\dot x)$ can not be inverted to give all the $\dot x^a(x,p)$.  
Given a set of primary constraints, $\phi_m$, the total Hamiltonian for the constrained system in phase space is
\[ H_T(q,p)=H(q,p) + u_m(q,p) \phi_m(q,p)\]
where $H(q,p)$ would be the Hamiltonian for the unconstrained system
and $u_m(q,p)$ are undetermined functions.

The time evolution of the primary constraints
can lead to more constraints, called \textit{secondary constraints}, $\phi_k$
with $k>\mathfrak M$,
and this process must be iterated until no more constraints are produced.
We may hit an inconsistency, \textit{e.g.} the Lagrangian $L=q$
  leads to an inconsistent equation of motion $0=1$. this is clearly
  nonsense and such systems must be rejected.  Assuming this does not happen
  there are three possibilities:

  \begin{itemize}
  \item Consistency of the time evolution gives the weak identity $0\approx 0$.

    \item Consistency constrains some (possibly all) of the $u$'s,
    $u_m\approx  U_m(q,p)$ with $ 0 \le m \le {\mathfrak M}$, where $U_m(q,p)$ are specific functions on phase space.

    \item Consistency generates new (secondary) constraints $\phi_k$.
    \end{itemize}
    Eventually the process terminates.

  Suppose there is a total of $\mathfrak J$ constraints on phase space $\phi_J(q,p)\approx 0$, $J=1,\ldots,\mathfrak{J}$, consisting of $\mathfrak M$ primary constraints and $\mathfrak K$ secondary constraints with $\mathfrak M + \mathfrak K=\mathfrak J$. 
Any function $R(q,p)$ on phase space that commutes weakly with all the constraints, in the sense that the Poisson brackets give $\{ R,\phi_I\}\approx 0$, is called \textit{first class}. Otherwise it is \textit{second class}.\footnote{The Hamiltonian for the relativistic particle in \S\ref{RelativisticParticle} is first class.}
If $R$ is a first class function its Poisson bracket with any of the constraints
is linear in the constraints,\footnote{Think of the Taylor expansion about $x=0$ of a function $f(x)$ that vanishes at $x=0$,
  \[f(x) = f(0) x + \frac 1 2 f'(0) x^2 + \cdots = r(x) x \]
for some function $r(x)$.}
\[ \{R,\phi_I\} = r_{I J} \phi_J.\]

\noindent \textbf{Theorem:} The Poisson bracket of two first class quantities is first class.

\medskip

\noindent \textbf{Proof:} Let
\[\{ R,\phi_I \} = r_{I J} \phi_J \qquad \mbox{and} \qquad \{S,\phi_I\} = s_{I J}, \phi_J\]
then, from the Jacobi identity,
\begin{align*} \{ \{R,S\},\phi_I\}&= -\{\{S,\phi_I\},R\} - \{\{\phi_I,R\},S\}\\
&  = -\{s_{I J} \phi_J,R\} + \{r_{I J}\phi_J,S\}\\
                                  & = -\{s_{I J},R\} \phi_J -s_{I J}\{\phi_J,R\}  + \{r_{I J},S\}\phi_J + r_{I J}\{\phi_J,S\}\\
                                  & = -\{s_{I J},R\} \phi_J + s_{I J} r_{J K}\phi_K  + \{r_{I J},S\}\phi_J
                                    - r_{I J} s_{J K}\phi_K\\
  &\approx 0.
\end{align*}

In particular any first class constraint $\phi_I$
satisfies
\[ \{\phi_I,\phi_J\}=c_{I J K}\phi_K\]
for all $J$ and some set of functions $c_{I J K}$.

 An important quantity for the subsequent analysis is the $\mathfrak J \times \mathfrak J$ anti-symmetric matrix
  \[ C_{I J} = \{ \phi_I, \phi_J\}.\]
Consider the time evolution of one of the constraints $\phi_I$, for any $I$ in the full range $1,\ldots,{\mathfrak J}$
\[ \dot \phi_I = \{\phi_I, H\} + u_m \{\phi_I, \phi_m\}
  = \{\phi_I, H\}  + C_{I m}u_m \approx 0.\]
Suppose we can a solution $u_m=U_m(q,p)$ of the inhomogeneous equations
\beq C_{I m} u_m  \approx \{H,\phi_I\},\label{inhomogeneous}\eeq
where the set of $u_m$ is viewed as the components of a single vector. 
We can add to this any non-trivial solution $V_m(q,p)$ of the homogeneous equation
\[ C_{I m} V_m =0,\]
and there may be a number of linearly independent such $V_m$'s, $V_{\alpha m}$, labelled by  $\alpha=1,\ldots,{\cal A}$.
\textit{i.e.} for each $\alpha$, $V_{\alpha m}$ is a non-trivial eigenvector of the matrix $C_{I J}$,
\[C_{I m} V_{\alpha m} =0 \]
with eigenvalue zero. If ${\cal A}>0$ then $C$ is not invertible and ${\cal A}$ is the number of
zero eigenvalues that it has. If ${\cal A}=0$ all the constraints are second class, $C$ is invertible and there are no $V$'s.

In general
\[ u_m \approx U_m(q,p) + v_\alpha V_{\alpha m}(q,p)\]
will be the most general solution of (\ref{inhomogeneous}), with the $v_\alpha$ undetermined functions.

To summarise, $u_m=U_m +  v_\alpha V_\alpha$ is the most general solution of
the algebraic matrix equation \[ \dot \phi_I  = \{\phi_I, H\}  + C_{I m}u_m \approx 0,\]
with $U_m$ a particular solution of the inhomogeneous equation
\[  C_{I m}U_m \approx  \{H,\phi_I\}\]
and $v_\alpha V_{\alpha m}$ the most general solution of the homogeneous equation
\[C_{I m} ( v_\alpha V_{\alpha m})=0,\]
where $v_\alpha$ are arbitrary functions.
The total Hamiltonian
\beq H_T = H +(U_m + v_\alpha V_{\alpha m}) \phi_m\label{TotalH}\eeq
has $\mathcal{A}$ undetermined functions $v_\alpha$ in it which are completely free and cannot affect the physics, they are associated with redundant gauge degrees of freedom in the description of the dynamics.
By definition only the primary constraints appear in the total Hamiltonian (\ref{TotalH}) but we saw in \S\ref{CPn} there is an undetermined function for every first class constraint, regardless of whether it is primary or secondary, so when there are first class constraints the most general Hamiltonian is the \textit{extended} Hamiltonian
\[ H_E = H _T+v_{\alpha'} \phi_{\alpha'}\]
where $\alpha'$ labels all first class secondary constraints (the first class primary constraints are already included in $H_T$) and $v_{\alpha'}$ are arbitrary functions. The time evolution of the constraints using $H_E$ is weakly the same as that using $H_T$, because the $\phi_{\alpha'}$ are first class. 
$H_E$ now contains all the primary constraints, both first and second class, and all the first class secondary constrains. We could extend the Hamiltonian even further and add the second class secondary constraints as $u_k \phi_k$, where $\phi_k$ are all the second class secondary constraints and the $u_k$ are initially  undetermined functions,
and define a completely extended Hamiltonian
\[H_{\cal E} = H_E + u_k \phi_k,\]
but $H_E$ has been constructed so that
\[ \dot \phi_J = \{ \phi_J,H_E\} \approx 0\]
  so 
  \[ \{ \phi_J,H_{\cal E}\} \approx u_k \{ \phi_J,\phi_k\}\]
  and, by definition of a secondary constraint, at least one of the $\{ \phi_J,\phi_k\}$ is not weakly zero for some $J$, so a consistent time evolution of the constraints using $H_{\cal E}$ requires $u_k \approx 0$. It makes no difference whether $H_E$ or $H_{\cal E}$ is used to calculate time evolution. At the end of the day the distinction between primary constraints and secondary constraints 
melts away and only the distinction between first class and second class constraints remains ---
first class constraints imply a redundancy in the degrees of freedom and a resulting ambiguity in the time evolution, exhibited in the arbitrary functions
$v_\alpha$ and $v_{\alpha'}$ in $H_{\cal E}$, which have no physical significance.

If there are no first class constraints ${\cal A}=0$ and $C$ is invertible.  We can then project the symplectic structure of the full phase space onto the reduced phases space using Dirac brackets,
\[ \{F,G\}_* = \{ F , G\} - \{F,\phi_I\} \bigl(C^{-1}\bigr)^{I J} \{ \phi_J,G\}.\]
While it is possible to  remove the redundancy in the phase space co-ordinates if ${\cal A}>0$
and there are first class constraints
(by choosing gauge conditions, increasing the number of constraints and increasing the size of $C$ in such a way as to remove the zero eigenvalues and render $C$ invertible)  this is not always useful.
It is often not possible to do this globally and different gauge choices must be made in different regions of phase space.

Finally note that none of the formal developments in this section assumed a finite number of degrees of freedom in the dynamical system.  Field theories have an infinite number of degrees of freedom, every field $\Psi$ has a degree of freedom associate with each point $\mathbf x$ in space, $\Psi(t,\mathbf x)=\Psi_{\mathbf x}(t)$, and ${\mathbf x}$ labels the degrees of freedom, just as the discrete index $a$ on $q^a(t)$ does for a system with a finite number of degrees of freedom.  Many relativistic field theories that give a good description of Nature,
such as electromagnetism, non-Abelian gauge theories and General Relativity, have Lagrangian descriptions that lead to Hamiltonian formulations with first class constraints. In the Lagrangian formulation of relativistic electromagnetism, for example, the fields are the four components $A_\mu$ of the relativistic vector potential and the time derivative of the electric scalar potential, $\dot A_0$, does not appear in the Lagrangian, so there is a constraint in phase space, that the momentum canonically conjugate to $A_0$ vanishes. Demanding consistency under time evolution shows that this is a first class constraint and Coulomb's law \[ \nabla.{\mathbf E} = \rho,\]
which is a dynamical equation in the Lagrangian formalism, emerges a secondary first class constraint in the Hamiltonian formalism, associated with $U(1)$ gauge invariance.
Not all four components of the vector potential $A_\mu$ are physical, there is a redundancy in the dynamical description that can be removed by a gauge choice. 
Similarly in Einstein's General Theory of Relativity the degrees of freedom are the 10 components $g_{\mu \nu}=g_{\nu\mu}$
of the space-time metric and the time derivative of the time-time component, $\dot g_{0 0}$,  does not appear in the Lagrangian and again there is a constraint in phase space, that the momentum canonically conjugate to $g_{0 0}$ vanishes. Again
consistency under time evolution shows that this is a first class constraint and leads to another first class constraint, that is essentially the time-time component of Einstein's equations, which is equivalent to the statement that the Hamilton itself vanishes, just as it does for the relativistic particle in \S\ref{RelativisticParticle}. The underlying symmetry giving rise to this first class constraint is diffeomorphism invariance of General Relativity and there is a redundancy in the dynamical description, not all ten components of the metric correspond to physical degrees of freedom of the theory.

\end{document}